**Superconductivity of Ta-Hf and Ta-Zr alloys: Potential alloys for use in superconducting devices**

Tomasz Klimczuk[1,2], Szymon Królak[1,2] and Robert J. Cava[3]

[1] *Faculty of Applied Physics and Mathematics, Gdansk University of Technology, ul. Narutowicza 11/12, 80-233 Gdańsk, Poland*

[2] *Advanced Materials Centre, Gdansk University of Technology, ul. Narutowicza 11/12, 80-233 Gdańsk, Poland,*

[3] *Department of Chemistry, Princeton University, Princeton, NJ 08544, USA*

tomasz.klimczuk@pg.edu.pl

rcava@princeton.edu

## Abstract

The electronic properties relevant to the superconductivity are reported for bulk Ta-Hf and Ta-Zr body centered cubic alloys, in large part to determine whether their properties are suitable for potential use in superconducting qbits. The body centered cubic unit cell sizes increase with increasing alloying. The results of magnetic susceptibility, electrical resistivity and heat capacity characterization are reported. While elemental Ta is a type I superconductor, the alloys are type II strong coupling superconductors. Although decreasing the electron count per atom is expected to increase the density of electronic states at the Fermi level and thus the superconducting transition temperature ($T_c$) in these systems, we find that this is not sufficient to explain the significant increases in the superconducting $T_c$'s observed.

## Introduction

Among the candidates for the fabrication of quantum computers, Qbit-based devices made from Josephson junctions enclosed in resonant cavities made from the superconductors Nb and Ta are of significant interest [1]. Of particular interest to us have been devices fabricated through the use of the elemental superconductor Ta in its body centered cubic (BCC) form [2]. This superconductor is type I, as most elements are, suggesting that it may be of interest to characterize devices made from Ta that has been modified. Devices made from doped BCC Ta may have improved properties since Ta is more noble than Nb, therefore restricting its weathering products, and finally that it may prove to be advantageous to increase the Tc of elemental Ta to higher temperatures (i.e higher than 4.4 K) to increase the temperature differential between the advent of the superconducting state and the temperature of device use. In this report we determine whether alloying the group V metal Ta with the group IV metals Hf and Zr has the chance to fulfill those criteria. In the bulk form, which is what is studied here, we find them to be advantageous in two of the three criteria. Previous work [3] has shown that $HfO_2$ is a very good dielectric material, the insulating nature of $ZrO_2$ is widely known, and both Hf and Zr are electropositive metals, and thus the alloys potentially fulfill all three of the hypothetical criteria for improved Qbit performance.

Several solid solutions of metals alloyed into the early transition metals, i.e. Ti-Mo [4], Ti-Rh [5], and of particular interest for this work into the group V metals have been synthesized and studied, i.e. Nb-Ti [6], Nb-Tc [7], Ta-Re [8], Ta-Hf, and Ta-W [9]. According to our knowledge, although the superconducting properties of the solid solutions of Ta with its neighboring transition metals have been reported [9], its superconducting properties when alloyed with Zr have been studied only on a Ta-Zr thin film produced by co-sputtering [10]. Here we provide a detailed analysis of the superconducting properties of the bulk Ta-Hf and Ta-Zr solid solutions. We find that both solid solutions, which decrease the effective electron count per metal atom from that of the pure group V element Ta, increase $T_c$ from about 4.4 K to about 6.7 K for $Ta_{0.8}Hf_{0.2}$ and to about 7.1 K for $Ta_{0.8}Zr_{0.2}$. All the doped phases of Ta fabricated by our methods are BCC in the composition range of interest, and we present clear evidence that the alloyed phases are type II superconductors, not type I superconductors as is elemental Ta. Finally, focusing on the materials physics of the two alloy systems, we point out that the composition-dependent $T_c$ behavior for the Ta-Hf and Ta-Zr alloys is not solely influenced by the increased density of electronic states (DOS), which is suggested by the apparent electron count variation of $T_c$, concluding that the Debye temperature of the alloys is at least as important and possibly the more important factor for determining $T_c$.

## Methods

Polycrystalline $Ta_{1-x}Hf_x$ and $Ta_{1-x}Zr_x$ samples were obtained by arc-melting stoichiometric Ta (slug, Alfa Aesar, 99.95%) with Hf (Alfa Aesar, 99.9%) and Zr (Onyxmet, Zr+Hf 99.9%), inside a high purity Ar-filled arc furnace (MAM-1 Edmund Buhler GmbH,) with Zr as a getter. Samples were melted four times and flipped upside down after each time to ensure homogeneity. The weight loss resulting from the melting procedure was negligible. The as-cast samples were used for the studies.

The crystal structures of the arc-melted samples were determined with a Bruker D2 Phaser 2nd generation X-ray diffractometer equipped with CuKα (λ = 1.5404 Å) and XE-T detectors. Due to the ductile nature of the synthesized compounds, instead of using powders, thin (~ 0.1 mm), flat plates were prepared, by rolling small pieces of the samples in a rolling mill (Durston DRM C 130). The experimental diffraction patterns were refined employing the LeBail method implemented in Bruker Topas software.

Characterization of the electronic properties relevant to the superconductivity was performed utilizing an Evercool II Quantum Design Physical Property Measurement System (PPMS). Magnetic properties were investigated with the vibrating sample magnetometer (VSM) option, using small pieces cut from the arc-melted buttons. The heat capacity was determined employing the two-τ method and subtracting the previously measured contribution from the vacuum grease (Apiezon N). For resistivity measurements, thin (~ 0.1 mm) plates initially prepared for X-ray diffraction were used. Each of them was cut into the desired shape, polished, and then 20 μm-diameter platinum wire leads were spark-welded to the sample surface.

## Results

Figure 1 presents a representative selection of the X-ray diffraction (XRD) patterns obtained on the flat thin cold rolled sheets. The black points represent the experimental data and the red solid line shows the profile refinement (LeBail) that was used to calculate the lattice parameters. For the pure Ta sample, the (110) XRD reflection (see right panel of Figure 1) is relatively broad with no traces of $K_{\alpha 1} - K_{\alpha 2}$ splitting, which is likely caused by the stress and defects induced by rolling. As Hf or Zr is added, the reflections become slightly broader. Our pure Ta metal forms in a BCC crystal structure with the reported lattice parameter $a$ = 3.3029 Å [11]. As is expected, since the Zr and Hf metallic radii are both larger than that of Ta, the refined lattice parameter $a$ linearly increases with $x$, in agreement with Vegard's law, (Figure 2). The initial slopes for both series (circles for $Ta_{1-x}Hf_x$ and squares for $Ta_{1-x}Zr_x$) are identical within error. (The d$a$(x)/dx's are 0.271(7) Å/mol and 0.270(9) Å/mol for the $Ta_{1-x}Hf_x$ and $Ta_{1-x}Zr_x$ series, respectively.) For Zr concentrations above 0.2, a deviation from linearity is observed, which we take as an indication of the Zr solubility limit in Ta. For larger Zr concentration the second phase detected by XRD studies is characterized by the hexagonal crystal structure of Zr metal. The expected Bragg reflections for the hexagonal phase (s.g. $P6_3/mmc$ #194) are represented by vertical blue bars in Figure 1(d) for $Ta_{0.5}Zr_{0.5}$. The refined lattice parameters are $a$ = 3.254(1) Å, $c$ = 5.086(2) Å. Whereas the $a$ is comparable, the $c$ is smaller than that reported for a pure Zr metal [12]. For the same Hf concentration (x = 0.5) the hexagonal phase is not yet present. In summary, alloying Zr or Hf with Ta causes an increase of the dimensions of the BCC unit cell, and the solubility limits of Zr and Hf in BCC Ta are determined to be 20-30% and 70-80%, respectively. All the data presented hereafter are only for the samples at or below the solubility limit.

The lower part of Figure 2 presents the superconducting critical temperature ($T_c$) as a function of Zr and Hf concentration in BCC $Ta_{1-x}M_x$. (The data points are taken from the heat capacity measurement, which will be discussed later in this manuscript.) For both dopants, in the low concentration range ($x \leq 0.2$), $T_c$ increases with increasing $x$. For a Hf concentration between 0.2 and 0.3, well within the BCC solid solution region a maximum $T_c$ is seen. One can expect similar behavior for the $Ta_{1-x}Zr_x$ system and in fact $T_c$(x) shows a tendency to flatten for x > 0.15 but the $T_c$ value for the samples with x(Zr) > 0.2 does not change (not presented here) up to 0.5 further confirming (as indicated by our XRD studies) that the Zr solubility limit in BCC Ta is ~0.2. It should be mentioned that slightly higher maximum $T_c$ = 7.5 K (for Zr concentration between 30% to 45%) was observed for Ta-Zr films prepared by co-sputtering [10]. This method likely extends the Zr concentration limit in the $Ta_{1-x}Zr_x$ alloy.

The inset of Figure 3 presents the temperature dependence of the normalized resistivity $\rho(T)/\rho(300K)$ for selected $Ta_{1-x}Hf_x$ samples (x = 0, 0.05, 0.1, 0.3, 0.5, 0.7). A similar figure for the $Ta_{1-x}Zr_x$ series is presented in Supplementary Material [13] (Fig. S1). Metallic-like behavior ($d\rho/dT > 0$) is seen for all the samples, but the RRR (residual resistivity ratio) decreases dramatically, from 35 (x = 0) to 1.1 (x = 0.7) – see main panel of Figure 3. The RRR parameter is a ratio, and is determined by $\rho(300K)/\rho_0$, where $\rho(300K)$ is the room temperature resistivity and $\rho_0$ is the resistivity at temperatures just above Tc. This parameter is often used to compare the quality of metallic compounds. All kinds of materials imperfections (e.g. vacancies, interstitials, grain boundaries, etc.) can cause scattering of charge carriers in addition to phonons. (Hence for a very good quality Ta sample prepared by the electrotransport technique at Ames Lab., the RRR obtained is 1700 [14], whereas for the polycrystalline Heusler-type superconductors, RRR is near 2 [15].) The rapid decrease in RRR for the $Ta_{1-x}Hf_x$ solid solution is likely to be mainly caused by the Hf(Zr) substitutional disorder, but the other effects, such as the possible decrease in the size of polycrystalline grains, may also be important. The RRR parameter for both dopants is almost identical.

The normalized resistivity at low temperatures, in the presence of various applied magnetic fields, is shown for the pure Ta, $Ta_{0.95}Hf_{0.05}$ and $Ta_{0.95}Zr_{0.05}$ in Figures 4(a) 4(b) and 4(c), respectively. A sharp transition to the superconducting state is seen at 4.47 K for Ta, 5.61 K for $Ta_{0.95}Hf_{0.05}$ and 5.83 K for $Ta_{0.95}Zr_{0.05}$. $T_c$, as is most commonly done for field-dependent transitions [16] is estimated as the midpoint of the resistivity transition, which is the intersection of the resistivity with the horizontal green line in the figures. As expected, $T_c$ shifts to lower temperature and the transition becomes broader as the applied magnetic field is increased.

Figure 4(d) presents the temperature dependence of the critical field ($H_c$) for a pure Ta sample (open circles) and the temperature dependence of the upper critical field ($H_{c2}$) for $Ta_{0.95}Hf_{0.05}$ (closed circles) and $Ta_{0.95}Zr_{0.05}$ (open squares). (In the following sections we will confirm type I superconductivity in Ta and type II superconductivity in the solid solution samples.) The solid lines presented in this figure are a fit to the expression

$$\mu_0 H_{c2}(T) = \mu_0 H_{c2}(0)\left[1-\left(\frac{T}{T_c}\right)^n\right] \quad (1).$$

The fit gives the zero Kelvin critical fields $\mu_0H_c(0)$ = 0.42(1) T for Ta, $\mu_0H_{c2}(0)$ = 1.57(8) T for $Ta_{0.95}Hf_{0.05}$ and $\mu_0H_{c2}(0)$ = 1.73(3) T for $Ta_{0.95}Zr_{0.05}$. The $\mu_0H_c(0)$ for pure Ta is much larger than the reported value of critical field (0.083 T) [17], thus suggesting the presence of filamentary regions, that display type II superconductivity. A similar effect was observed in elemental rhenium, in which the shear strain causes changes in the M(H) shape and almost double the critical temperature [18]. The estimated $\mu_0H_{c2}(0)$ values for the whole series of $Ta_{1-x}M_x$ are presented in Figure 6(b). The largest value of $\mu_0H_{c2}$ = 11.9(3) T is seen for $Ta_{0.5}Hf_{0.5}$. The *n* parameter of the fit changes from 1 to 1.34. A nearly linear $H_{c2}(T)$ has been previously reported for $WB_{4.2}$ [19], as well as for several iron based superconductors [20,21] and $Nb_2Pd_{0.81}S_5$ [22].

The next figure presents the magnetic measurements. The temperature dependence of the zero field cooled (ZFC) and field cooled (FC) volume magnetic susceptibility ($\chi_V = M_V/H$) in a field of H = 10 Oe is presented in Figure 5(a), 5(b) and 5(c) for Ta, $Ta_{0.95}Hf_{0.05}$ and $Ta_{0.95}Zr_{0.05}$, respectively. (The data are normalized by $1/4\pi$ and are also corrected for the diamagnetization factor N. At the lowest temperature, the superconducting signal exceeds that expected (in the CGS unit system) suggesting a full Meissner state in both samples, and we therefore take $\chi_V = -1/4\pi(1-N)$.) Compared with the ZFC data, the FC magnetic susceptibility signal is roughly 15% weaker for Ta and 90% weaker for $Ta_{0.95}Hf_{0.05}$. Since this measurement was performed in very low field (10 Oe), the flux pinning effect in the mixed state is unlikely. The other scenario assumes that the magnetic flux is trapped during the FC cooling at the grain boundaries. Comparing the FC signals strength of Ta and $Ta_{0.95}M_{0.05}$ (M = Hf, Zr) we can conclude that our pure Ta can be characterized as an almost single crystal. While the ZFC Meissner fraction stays constant at 100%, in the $Ta_{1-x}Hf_x$ solid solution, as the Hf concentration increases the relative ratio of the FC to ZFC magnetic susceptibility decreases from 10% for $Ta_{0.95}Hf_{0.05}$ to 1% for $Ta_{0.3}Hf_{0.7}$ (see Figure S1 in the Supplementary Material [13]).

Figures 5(d)-5(f) present the volume magnetization ($M_V$) vs. magnetic field (H), measured at constant temperature, below $T_c$, for elemental Ta and the $Ta_{.95}Hf_{.05}$ and $Ta_{.95}Zr_{.05}$ alloys. The measurements show a linear response with the minimum value between 430 Oe and 450 Oe. For elemental Ta (panel (d)), the shape of $M_V(H)$ suggests type I superconducting behavior. The lack of a rapid drop below the minimum of $M_V(H)$ is caused by a demagnetization effect. In contrast, type II superconductivity is suggested by the $M_V(H)$ curves for the $Ta_{0.95}Hf_{0.05}$ and $Ta_{0.95}Zr_{0.05}$ samples. Assuming a full Meissner effect, the initial slope of $M_V(H)$ can be used to estimate the demagnetization factor N, $4\pi\chi_V = 4\pi M_V/H = -1/(1-N)$. For Ta, $Ta_{0.95}Hf_{0.05}$ and $Ta_{0.95}Zr_{0.05}$ the N values are 0.44, 0.37 and 0.57 respectively.

The critical field ($H_c^*$) for the Ta sample was calculated at each temperature, as shown in panel (d), and $H_c^*(1.9\ K)$ is marked by an arrow. (The same method was used for obtaining the critical field for $CaBi_2$ [23].) The shape of $M_V(H)$ for $Ta_{0.95}Hf_{0.05}$ suggests type II superconductivity. The lower critical field ($H_{c1}^*$) is the field at which the first deviation from the Meissner state is observed (see the vertical arrow in panels (e) and (f)). In order to estimate $H_{c1}^*$, we used the method described in ref. [24]; the obtained values are presented in the panels (h) and (i) for $Ta_{0.95}Hf_{0.05}$ and $Ta_{0.95}Zr_{0.05}$, respectively. The extrapolated to 0 K values of $H_c^*(0)$ and $H_{c1}^*(0)$ are the fitting parameters in the widely used formula for the critical field (type-I) and the lower critical field (type-II):

$$H_{c(1)}^*(T) = H_{c(1)}^*(0)\left[1 - \left(\frac{T}{T_c}\right)^2\right] \quad (2).$$

The fits give 920(12) Oe and 247(6) Oe and hence correcting by a demagnetization factor ($H_{c(1)} = H_{c(1)}^{*}/(1-N)$), we obtained $H_c(0)$ = 1650 Oe, $H_{c1}(0)$ = 394 Oe and $H_{c1}(0)$ = 332 Oe for pure Ta, $Ta_{0.95}Hf_{0.05}$ and $Ta_{0.95}Zr_{0.05}$, respectively. The estimated critical field value for elemental Ta is larger than reported in ref. [17] and will be discussed later. The $\mu_0H_{c1}(0)$ values for the whole series of $Ta_{1-x}M_x$ are plotted in Figure 6(a) in the SI units (mT) for easier comparison with other critical fields shown in panel (b) and panel (c). Since there are only four points, which are scattered, it is difficult to comment on a tendency for the Zr-series. For the Hf-series we conclude that as the Hf concentration increases the lower critical field decreases by a factor of 2.

Knowing the lower and the upper critical fields, several additional superconducting properties can be calculated. From the Ginzburg-Landau formula for $H_{c2}$:

$$H_{c2}(0) = \frac{\Phi_0}{2\pi\xi_{GL}^2} \qquad (3)$$

we can obtain the superconducting coherence length $\xi_{GL}$. In the above formula $\Phi_0 = hc/2e$ is the flux quantum. Having $\xi_{GL}$ and $H_{c1}$, one can numerically calculate the superconducting penetration depth $\lambda_{GL}$ from the formula for the lower critical field:

$$H_{c1}(0) = \frac{\Phi_0}{4\pi\lambda_{GL}^2}\ln\frac{\lambda_{GL}}{\xi_{GL}} \qquad (4).$$

The next parameter that relates the calculated coherence length and penetration depth is the Ginzburg-Landau parameter $\kappa_{GL} = \lambda_{GL}/\xi_{GL}$.

The estimated values of $\lambda_{GL}$, $\xi_{GL}$ and $\kappa_{GL}$ for the $Ta_{1-x}Hf_x$ and $Ta_{1-x}Zr_x$ alloys are presented in Figure 7. The penetration depth increases with the dopant concentration from ~ 900 Å to ~ 1500 Å, whereas in contrast the coherence length decreases almost 3 times as *x* increases from 0.05 to 0.2 and then becomes *x* independent at $\xi_{GL}$ ~ 50 Å for the Hf content x > 0.2. The lowest value of $\kappa_{GL}$ is obtained for the lowest concentration of Hf and Zr substitution, $\kappa_{GL}$ = 5.9 and 7.2 respectively, which are both larger than $1/\sqrt{2}$, thus confirming type II superconductivity for both $Ta_{0.95}Zr_{0.05}$ and $Ta_{0.95}Hf_{0.05}$. Increasing the substitution level results in an increase in the $\kappa_{GL}$ parameter and saturation at a value ~ 30. Finally, using the formula:

$$H_{c1}H_{c2} = H_c^2 \ln \kappa_{GL} \qquad (5)$$

we can obtain the thermodynamic critical field $H_c(0)$. These values are plotted in Figure 6(c). Although the points are rather scattered, one can conclude that the thermodynamic critical field increases from $\mu_0H_c(0)$ = 0.165 T (pure Ta) and reaches the maximum at 0.37 T for the Hf concentration x = 0.4.

The final method that we used to characterize superconducting properties of $Ta_{1-x}M_x$ is the low heat capacity at temperatures near those of Tc. The zero-applied-field temperature dependent $C_p/T$ vales for Ta (open circles), $Ta_{0.95}Hf_{0.05}$ (closed circles) and $Ta_{0.95}Zr_{0.05}$ (open squares) are shown in a Figure 8 (a), (c) and (e), respectively. The bulk superconducting transition temperatures, estimated by using the equal entropy construction method (solid lines), are thus $T_c$ = 4.45 K, 5.21 K and 5.31 K for pure Ta and the 5% substituted Ta-Hf and Ta-Zr alloys, respectively. Although for the Ta sample the critical temperature is almost the same as obtained from the other techniques, this is not the case for the doped samples. The higher $T_c$ value observed by $\rho(T)$ and $\chi(T)$ is caused by the surface superconductivity and has been observed for many compounds (see for example ref. [25]). Panels (b), (d) and (f) present the $C_p/T$ vs. $T^2$ data under a 3 T applied magnetic field. This field exceeds the upper critical field for both tested samples, and hence the normal state can be fitted using the equation $C_p/T = \gamma + \beta T^2 + \delta T^4$. The first term here represents an electron contribution ($\gamma T$) and the last two terms are phonon contributions to the specific heat. The fits are shown as the solid lines and the values $\gamma$ = 6.02(9) mJ mol$^{-1}$ K$^{-2}$, $\beta$ = 0.110(8) mJ mol$^{-1}$ K$^{-4}$, $\delta$ = 0.4(1) μJ mol$^{-1}$ K$^{-6}$ are obtained for Ta, the values $\gamma$ = 6.64(11) mJ mol$^{-1}$ K$^{-2}$, $\beta$ = 0.110(13) mJ mol$^{-1}$ K$^{-4}$, $\delta$ = 0.8(3) μJ mol$^{-1}$ K$^{-6}$ and $\gamma$ = 6.64(14) mJ mol$^{-1}$ K$^{-2}$, $\beta$ = 0.107(16) mJ mol$^{-1}$ K$^{-4}$, $\delta$ = 0.6(3) μJ mol$^{-1}$ K$^{-6}$ are obtained for $Ta_{0.95}Hf_{0.05}$ and $Ta_{0.95}Zr_{0.05}$, respectively. Knowing $\beta$ one can calculate the Debye temperature by using the expression:

$$\theta_D = \left(\frac{12\pi^4 Rn}{5\beta}\right)^{1/3} \quad (6),$$

where n = 1 (the number of atoms per f.u.) and R = 8.31 J mol$^{-1}$. The Debye temperatures thus obtained for Ta and for the $Ta_{0.95}Hf_{0.05}$ and $Ta_{0.95}Zr_{0.05}$ alloys are $\Theta_D$ = 256(6) K and 260(10) K, 263(13) K. The one we obtain for elemental Ta (256 K), is very close to the one reported in the literature for pure Ta metal (246 K) [26].

Having the Sommerfeld coefficient ($\gamma$), an important superconducting parameter can be calculated, ($\Delta C/\gamma T_c$). The Bardeen-Cooper-Schrieffer theory predicts that for a weakly-coupled superconductors, the normalized specific heat jump, $\Delta C/\gamma T_c$ = 1.43. The values we obtained are 1.53 (Ta), 1.71 ($Ta_{0.95}Hf_{0.05}$), 1.70 ($Ta_{0.95}Zr_{0.05}$) and suggest moderately strong coupling superconductivity for the alloys. Figure 9 gathers the Sommerfeld coefficient, the Debye temperature and the normalized solid solution superconducting jump in the heat capacity for for the $Ta_{1-x}Hf_x$ and $Ta_{1-x}Zr_x$ alloy series. As *x* increases, $\gamma$ increases, reaching a maximum value of $\gamma$ = 7.82(16) mJ mol$^{-1}$ K$^{-2}$ for x(Hf) = 0.2. Although this is a relatively large value, the Sommerfeld parameter for the $Nb_{1-x}Zr_x$ and $V_{1-x}Ti_x$ alloys is even larger, reaching ~11 mJ mol$^{-1}$ K$^{-2}$ [27].

Having the Sommerfeld parameter and critical temperature we can calculate the thermodynamic critical field by using the equation provided by the $\alpha$-model [28,29]:

$$\frac{H_c(0)}{(\gamma T_c^2)^{1/2}} = \sqrt{\frac{6}{\pi}}\,\alpha \qquad (7)$$

In this equation the $\alpha$ parameter was obtained from the following relation:

$$\frac{\Delta C}{\gamma T_c} \approx 1.426\left(\frac{\alpha}{\alpha_{BCS}}\right)^2 \qquad (8)$$

where $\alpha_{BCS}$ = 1.764. For pure Ta we obtained $\alpha$ = 1.806 and taking $\gamma$ = 5800 erg cm$^{-3}$ K$^{-2}$ (in eq. 7 CGS units are used) the thermodynamic critical field is $H_c$ = 837 Oe which is in perfect agreement with the critical field value (830(4) Oe) provided from the precise ballistic-induction measurement [17].

Interesting behavior is observed for the composition dependence of the Debye temperature. Considering the uncertainty in the estimation of $\Theta_D$, one can conclude that for x < 0.2 it does not change in the Hf series and slightly decreases in the Zr-series. For x(Hf) > 0.2, the Debye temperature decreases by 40% and flattens at around 155 K Hafnium concentrations *x*(Hf) > 0.5. Since the molar mass of Zr and Hf are smaller than the molar mass of Ta, and thus the Debye temperature is expected to decrease, the behavior shown in Figure 9(b) cannot be explained by a simple monoatomic model with decreasing atomic mass. The obtained values are in agreement with those reported earlier, but it should be noted that the authors provide $\Theta_D$ only for Ta, $Ta_{0.7}Hf_{0.3}$ and $Ta_{0.39}Hf_{0.61}$.

The final figure presents $\Delta C/\gamma T_c$ for both of the $Ta_{1-x}M_x$ series. The heat capacity jump increases monotonically from $\Delta C/\gamma T_c$ = 1.53 (Ta) to $\Delta C/\gamma T_c$ = 2.2 for $Ta_{0.2}Hf_{0.8}$. This large value (significantly different from the BCS weak coupling superconductor value of 1.43) is comparable to the values reported for the strong coupling superconductors Pb (2.66) [30] and Pb-Bi alloys (2.9-3.0) [31].

Figures 10(a)-(c) present the temperature dependence of $C_p/T$ under zero and low magnetic field in the vicinity of the superconducting transition. The data were collected by using a single, large heat pulse ($\Delta T$ = 80% of the base temperature) and re-processed by using the single-slope post processing method provided by the PPMS MultiVu package (Quantum Design). Obviously different behavior is observed for Ta (a) and the $Ta_{0.95}Hf_{0.05}$ (b) and $Ta_{0.95}Zr_{0.05}$ (c) alloys. Whereas for $Ta_{0.95}Hf_{0.05}$ and $Ta_{0.95}Zr_{0.05}$ a suppression of the $\Delta C/T_c$ is observed, this is not a case for pure BCC Ta, for which the $\Delta C/T_c$ initially increases and then decreases with applied magnetic field. This behavior is an indication of the crossover from a 2$^{nd}$ to a 1$^{st}$ order phase transition and proves [32] that Ta is a type-I superconductor. The absence of this effect for $Ta_{0.95}Hf_{0.05}$ and $Ta_{1-x}Zr_x$ samples means that the alloying causes a transition from type-I to type-II superconductivity. Experimental data obtained by this method with a magnetic field step of 20 Oe. allows us to plot the contour map of $C_p/T$ as a function of temperature and applied magnetic field. A similar analysis (Figure S3) is provided for pure Nb metal in the Supplementary Material [13]. Nb is a rare example of an element in which type-II superconductivity is observed.

The next superconducting parameter that can be obtained is the electron-phonon coupling constant ($\lambda_{ep}$), typically calculated by using the inverted McMillan formula [33]. However, since the $\Delta C/\gamma T_c$ value is large and increases with the Hf or Zr content of the alloy (for x = 0.5, $\Delta C/\gamma T_c = 2.2$), it is likely that alloying shifts the system from moderately to strongly coupled superconductivity. Hence, the Allen-Dynes formula [34] should be used:

$$T_c = \left(\frac{\omega_{ln}}{1.2}\right) exp\left[-\frac{1.04(1+\lambda_{ep})}{\lambda_{ep}-\mu^*(1+0.62\lambda_{ep})}\right] \quad (9),$$

where $\mu^*$ is the Coulomb repulsion parameter and $\omega_{ln}$ is the logarithmically averaged phonon frequency, which can be determined from the specific heat capacity jump at $T_c$:

$$\left[\frac{\Delta C_p}{\gamma T_c}\right]_{T_c} = 1.43\left[1+53\left(\frac{T_c}{\omega_{ln}}\right)^2 ln\left(\frac{\omega_{ln}}{3T_c}\right)\right] \quad (10).$$

Taking $\mu^* = 0.13$, we obtain $\lambda_{ep} = 0.65, 0.84$ and $0.83$ for Ta, $Ta_{0.95}Hf_{0.05}$ and $Ta_{0.95}Zr_{0.05}$, respectively, consistent with the large $\Delta C/\gamma T_c$ values deduced from the alternative analysis. Finally, knowing the $\lambda_{ep}$ and $\gamma$ values, the electron density of electronic states at the Fermi energy $DOS(E_F)$ can be calculated using the relation [35]:

$$DOS(E_F) = \frac{3\gamma}{\pi^2 k_B^2(1+\lambda_{el})} \quad (11).$$

For both the $Ta_{1-x}Hf_x$ and $Ta_{1-x}Zr_x$ alloys, the thus determined $\omega_{ln}$, $\lambda_{el}$ and $DOS(E_F)$ values together with the measured superconducting critical temperature are presented in Figure 11. (Note that instead of *x*, the valence electron number per atom ($N_{el.}$/at.) is used in this Figure. Allowing us to compare our results directly with those reported in refs. [9] and [27].)

The Allen-Dynes formula [34] for the superconducting transition temperature includes the logarithmically averaged phonon frequency and the electron – phonon coupling parameter.

As we decrease the valence electron number (increase of the Hf or Zr alloying element content) $\lambda_{el}$ increases whereas $\omega_{ln}$ stays constant or decreases in the whole doping range. The consequence of the opposing behavior of the two parameters that combine to yield the superconducting transition temperature for the $Ta_{1-x}Hf_x$ series is the observed maximum of the critical temperature, which occurs for $N_{el.}$ = 4.7 el./at. This is not the value for which the maximum of DOS($E_F$) is seen ($N_{el.}$ ~ 4.9 el./at.).

**Conclusion**

In summary, we have synthesized and explored the superconducting characteristics of the two solid solution systems $Ta_{1-x}Hf_x$ and $Ta_{1-x}Zr_x$ in bulk form. The latter system has not been reported in the past. The refined cubic lattice parameter *a* of the BCC phases formed increases linearly with Zr and Hf concentration at low substitution levels, and the increases are almost identical. The solubility limit for Hf (between $x$ = 0.7 and 0.8) is much larger than for Zr (between $x$ = 0.2 and 0.3). The transition temperature to the superconducting state increases for both systems. Our magnetization and specific heat capacity measurements confirm that pure Ta is a type-I superconductor and further show that the lowest degree of Zr or Hf alloying (0.05) causes a transition of the Ta to type-II superconductivity. The lower critical field for the doped $Ta_{1-x}M_x$ alloys shows a decrease of $H_{c1}$, whereas the upper critical field reveals a maximum of $H_{c2}$ for x = 0.5. Similar behavior is observed for the thermodynamic critical field, but the maximum occurs at lower Hf concentration ($x$ = 0.4).

The Sommerfeld parameter, calculated based on the specific heat capacity results, initially increases and reaches maximum for the Hf doping level of $x$ = 0.2. Our most intriguing results, we argue, concern the behavior of the Debye temperature of the alloys. A small level of alloying does not affect the Debye temperature, and then a decrease in $\Theta_D$ is observed for Hf concentrations of $0.2 < x < 0.5$, and finally for $x$ above 0.5 the Debye temperature reaches a constant value of about 155 K. This is unexpected behavior and should be verified by other experiments, as well as by phonon density of states calculations for the Ta-Hf solid solution.

The observed change of the calculated electron-phonon coupling value as well as the normalized specific heat jump at the superconducting transition suggest that the $Ta_{1-x}M_x$ systems studied here change from moderately to strongly coupled superconductivity with increasing alloying. The observed $\Delta C/\gamma T_c$ = 2.2 value for $Ta_{0.3}Hf_{0.7}$ is very large and comparable to the values reported for Pb and Pb-Bi alloys, which are known to be strong coupling superconductors. Our results confirm a maximum of $T_c$ for $N_{el}$ between 4.7 and 4.8 electrons/atom [7] in the $Ta_{1-x}Hf_x$ system. This behavior cannot be explained by the change in the DOS($E_F$) alone, which is why we emphasize that the logarithmically averaged phonon frequency is an important parameter to explain the behavior of the critical temperature for the $Ta_{1-x}Hf_x$ solid solution system. Finally, we conclude that the superconducting characteristics of the Ta-Hf and Ta-Zr alloy systems, especially at the alloying levels of 5 to 20%, make them worthy of study in thin film form in superconducting Qbit-based devices.

## Acknowledgments

This report is based upon work supported by the U.S. Department of Energy, Office of Science, National Quantum Information Science Research Centers, Co-design Center for Quantum Advantage (C2QA) under contract number DE-SC0012704. The work at Gdansk Tech. was supported by the National Science Centre (Poland; Grant UMO-2018/30/M/ST5/00773). Sz. K. acknowledges the Excellence Initiative - Research University, project 11/RADIUM/2022. We kindly acknowledge discussions with dr Bartlomiej Wiendlocha and dr Sylwia Gutowska from AGH (Cracow, Poland).

## Figures

**Figure 1** The powder X-ray diffraction patterns for pure Ta (a), $Ta_{0.9}Zr_{0.1}$ (b), $Ta_{0.9}Hf_{0.1}$ (c), $Ta_{0.5}Zr_{0.5}$ (d) and $Ta_{0.5}Hf_{0.5}$ (e). Experimental data points are represented by open circles, whereas a fitting line (profile refinement) is shown by a red line. The expected Bragg reflections for the cubic bcc (Im-3m) and for the hexagonal $P6_3/mmc$ crystal structure are represented by red and blue vertical bars, respectively. The right panel presents the XRD patterns focused on the (110 ) reflection.

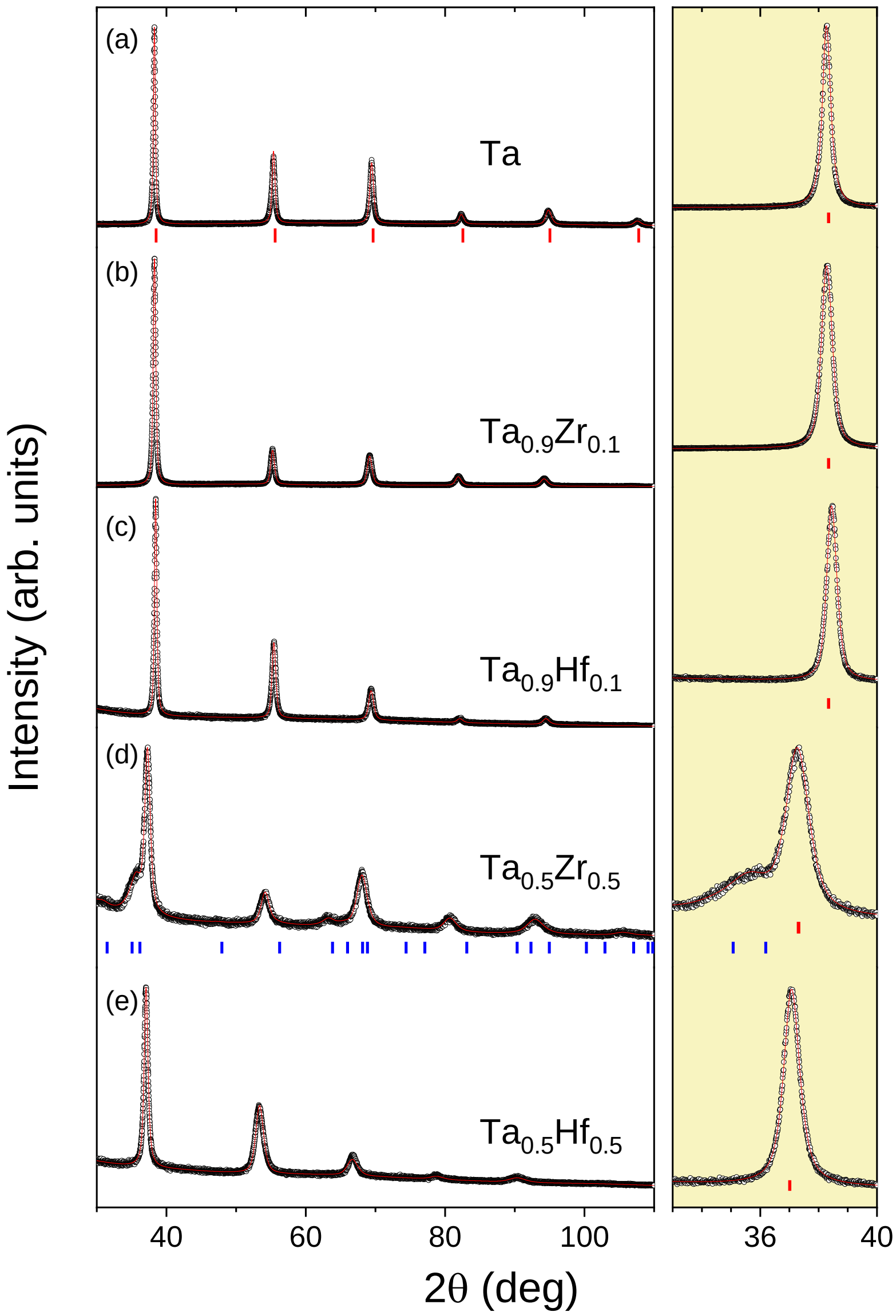

**Figure 2** The lattice parameter and $T_c$ as a function of the concentration of Hf (open circles) and Zr (squares) in $Ta_{1-x}M_x$ solid solutions. The solubility limit for Zr was revealed to be 0.2, while for Hf it reaches 0.7.

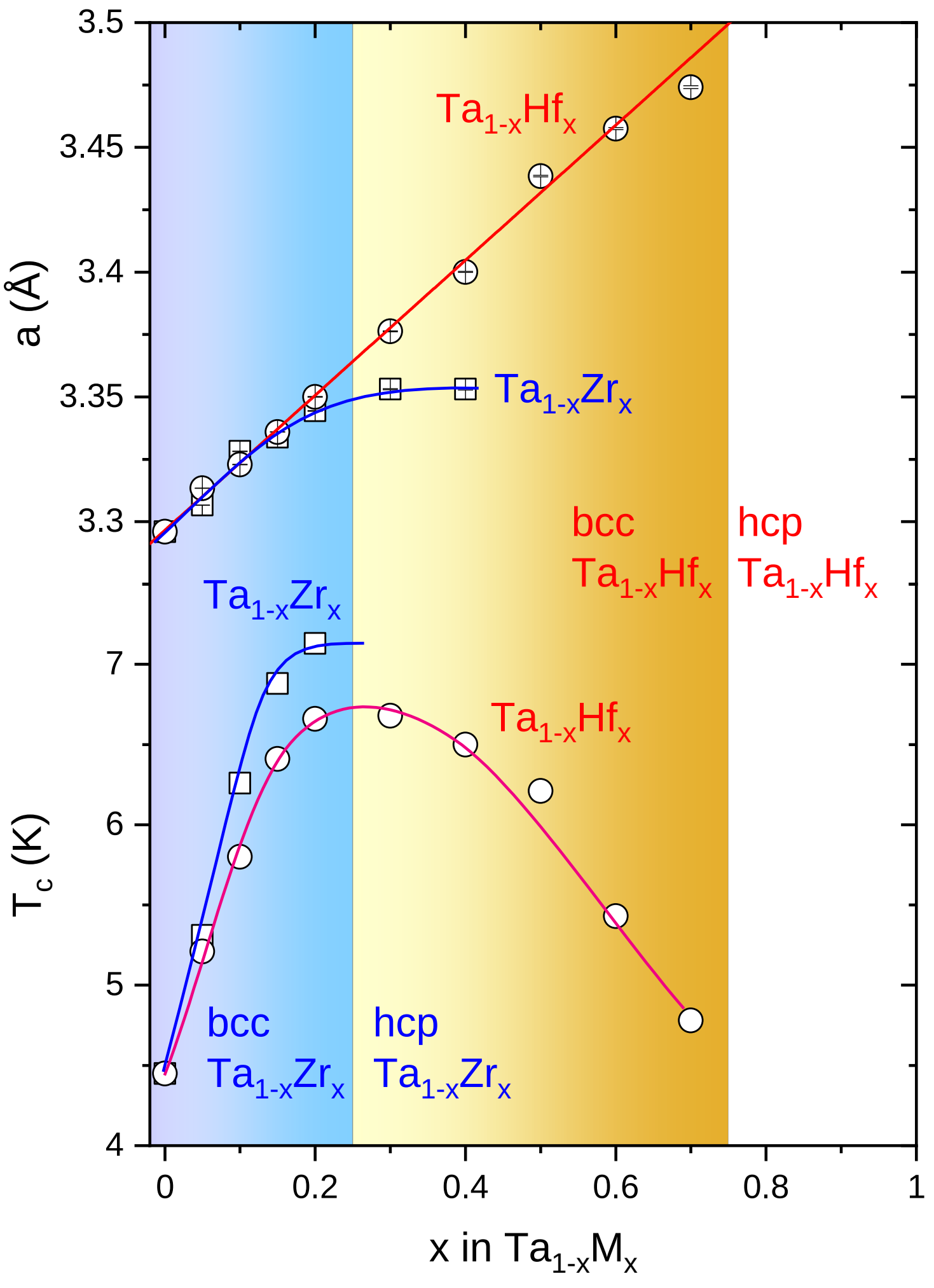

**Figure 3** Concentration dependence of the residual-resistivity-ratio (RRR) for $Ta_{1-x}Hf_x$ (closed circles) and $Ta_{1-x}Zr_x$ (open squares). The inset shows temperature dependence of the normalized electrical resistivity ρ(T)/ρ(300 K) for selected $Ta_{1-x}Hf_x$ samples (x = 0, 0.05, 0.1, 03, 0.5 and 0.7).

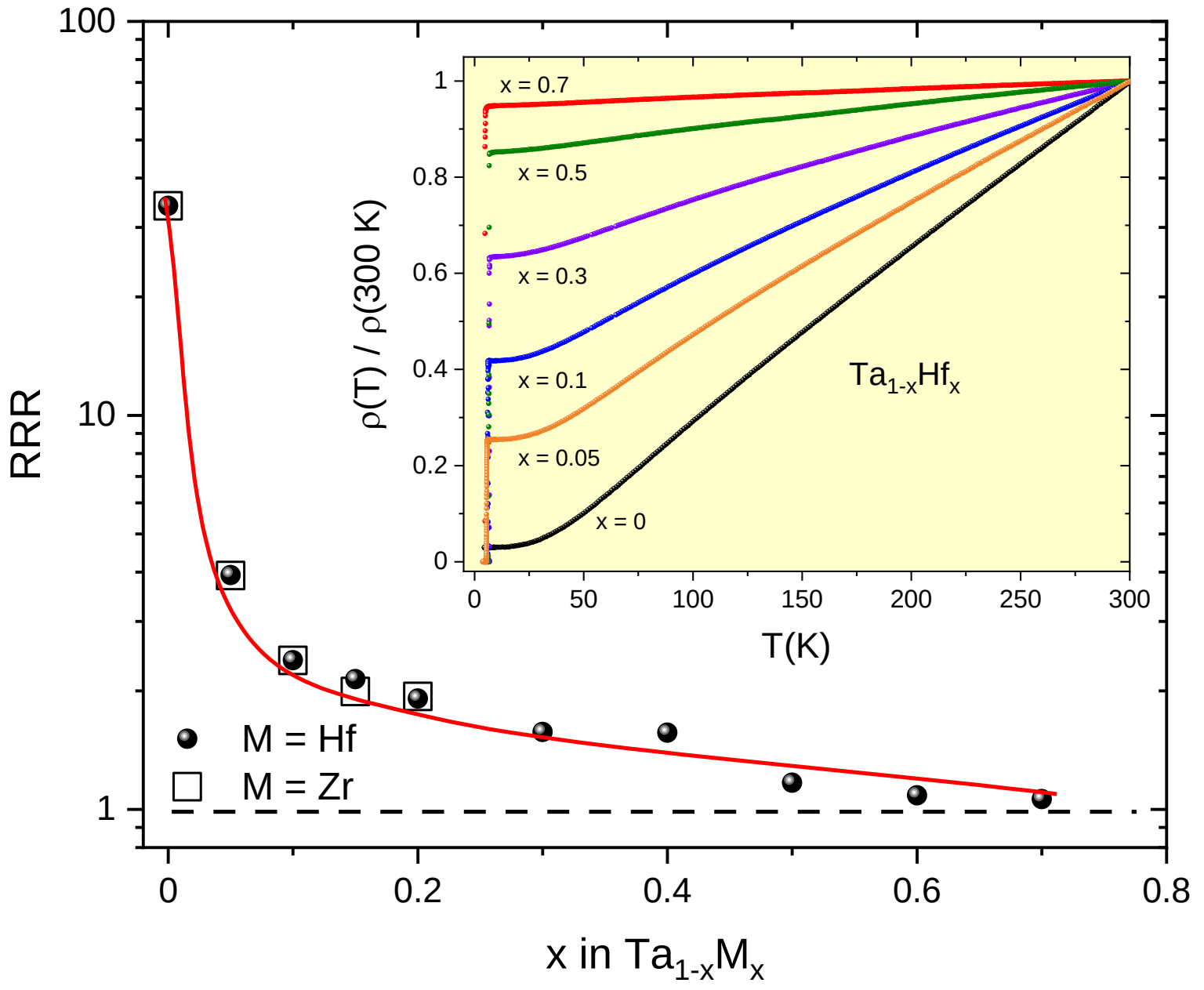

**Figure 4** Temperature dependence of the normalized electrical resistivity ρ(T)/ρ(300K) for Ta (a), $Ta_{0.95}Hf_{0.05}$ (b) and $Ta_{0.95}Zr_{0.05}$ (c) measured in zero field and under various applied magnetic field $\mu_0H$ as indicated. Panel (d) presents the critical field ($\mu_0H_c$) and the upper critical field ($\mu_0H_{c2}$) vs. temperature for Ta, $Ta_{0.95}Hf_{0.05}$ and $Ta_{0.95}Zr_{0.05}$, respectively.

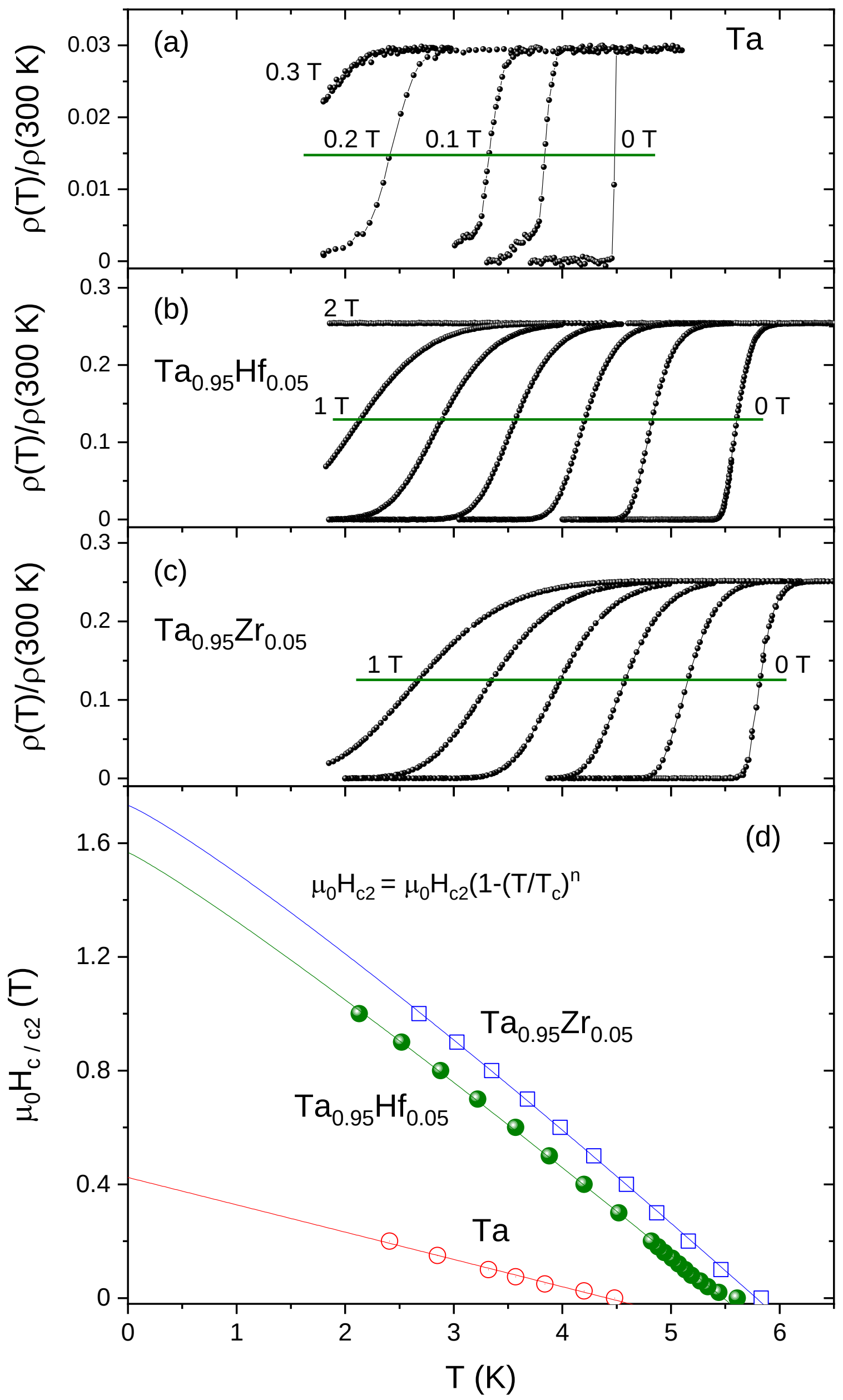

**Figure 5** Magnetic data for Ta and $Ta_{0.95}Hf_{0.05}$. Panel (a), (b) and (c) – normalized field-cooling (FC) and zero-field-cooling (ZFC) volume magnetic susceptibility measured under low magnetic field (H = 10 Oe). Panel (d) to (f) – volume magnetization vs. magnetic field measured at constant temperatures below $T_c$. The critical field ($H_c$*) and the lower critical field ($H_{c1}$*) for Ta, $Ta_{0.95}Hf_{0.05}$ and $Ta_{0.95}Zr_{0.05}$ are presented in panel (g), (h) and (i), respectively.

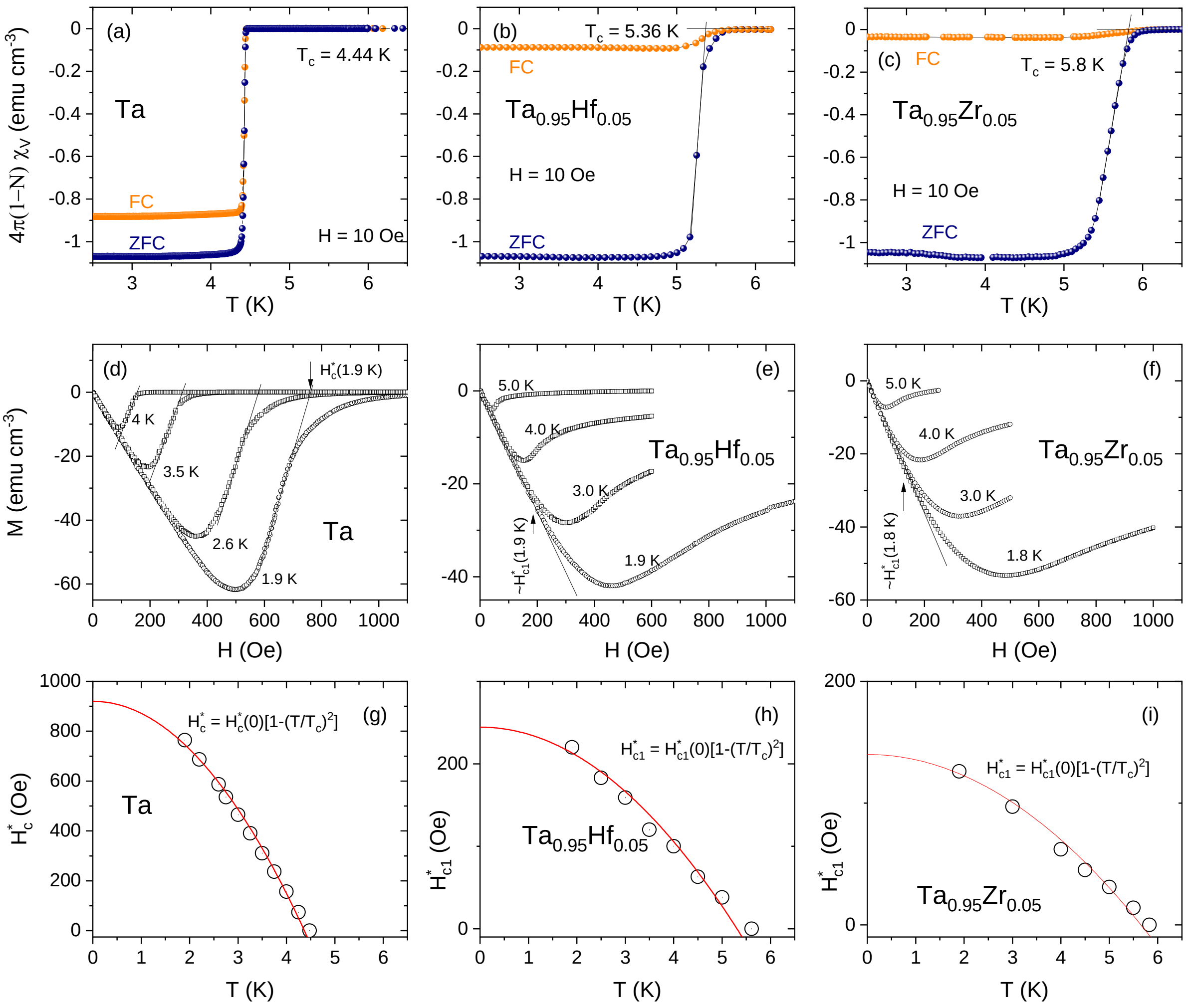

**Figure 6** The concentration dependences of the lower critical field $\mu_0H_{c1}$ (a), upper critical field $\mu_0H_{c2}$ (b) and thermodynamic critical field $\mu_0H_c$. Note that, since Ta is a type-I superconductor, only $\mu_0H_c$ is marked. The red solid lines are guide lines.

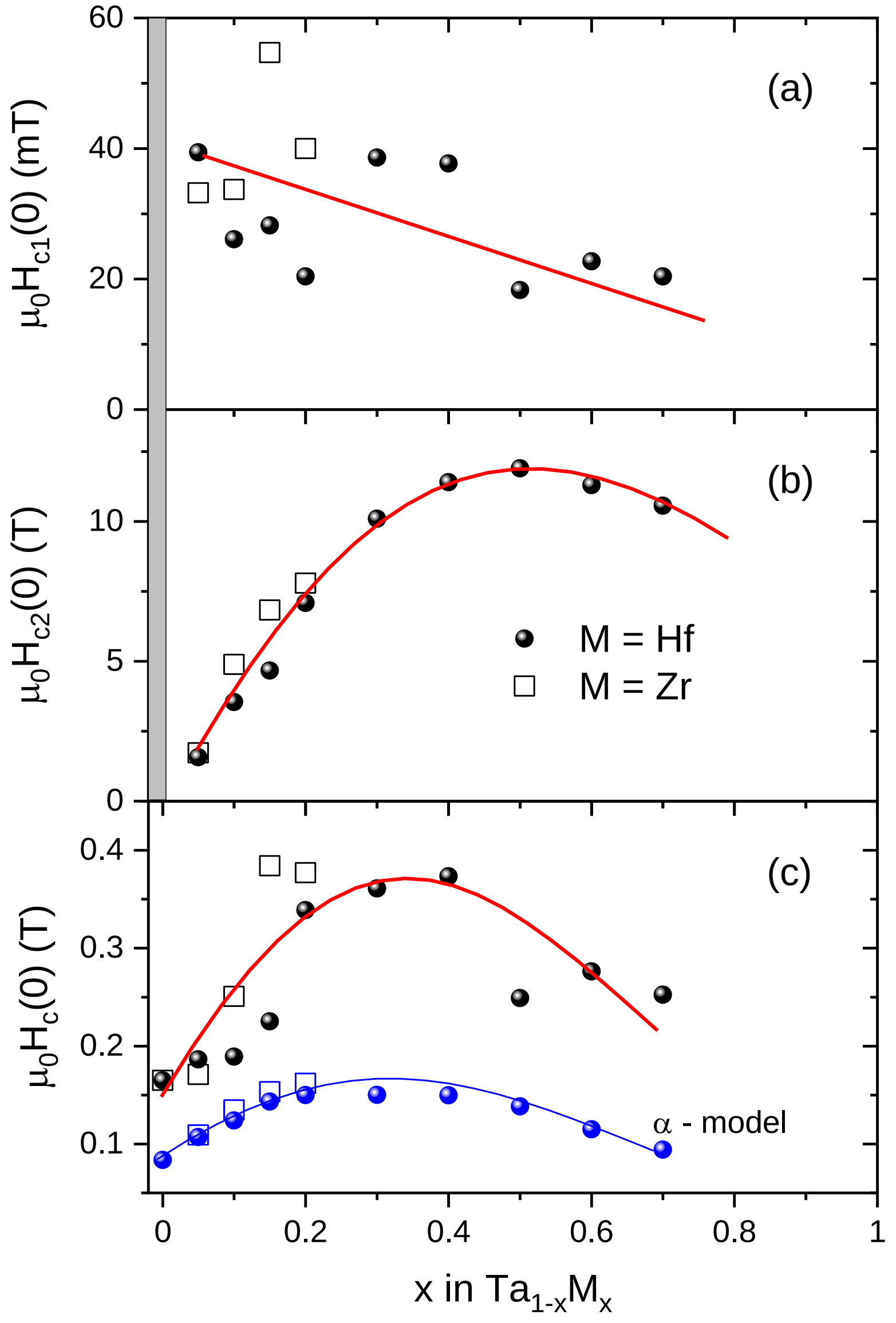

**Figure 7** Panel: (a) penetration depth (λ), (b) coherence length (ξ) and (c) the Ginzburg-Landau parameter κ vs. concentration of the M atom in $Ta_{1-x}M_x$ solid solution (M = Hf and Zr). The red solid lines are guide lines.

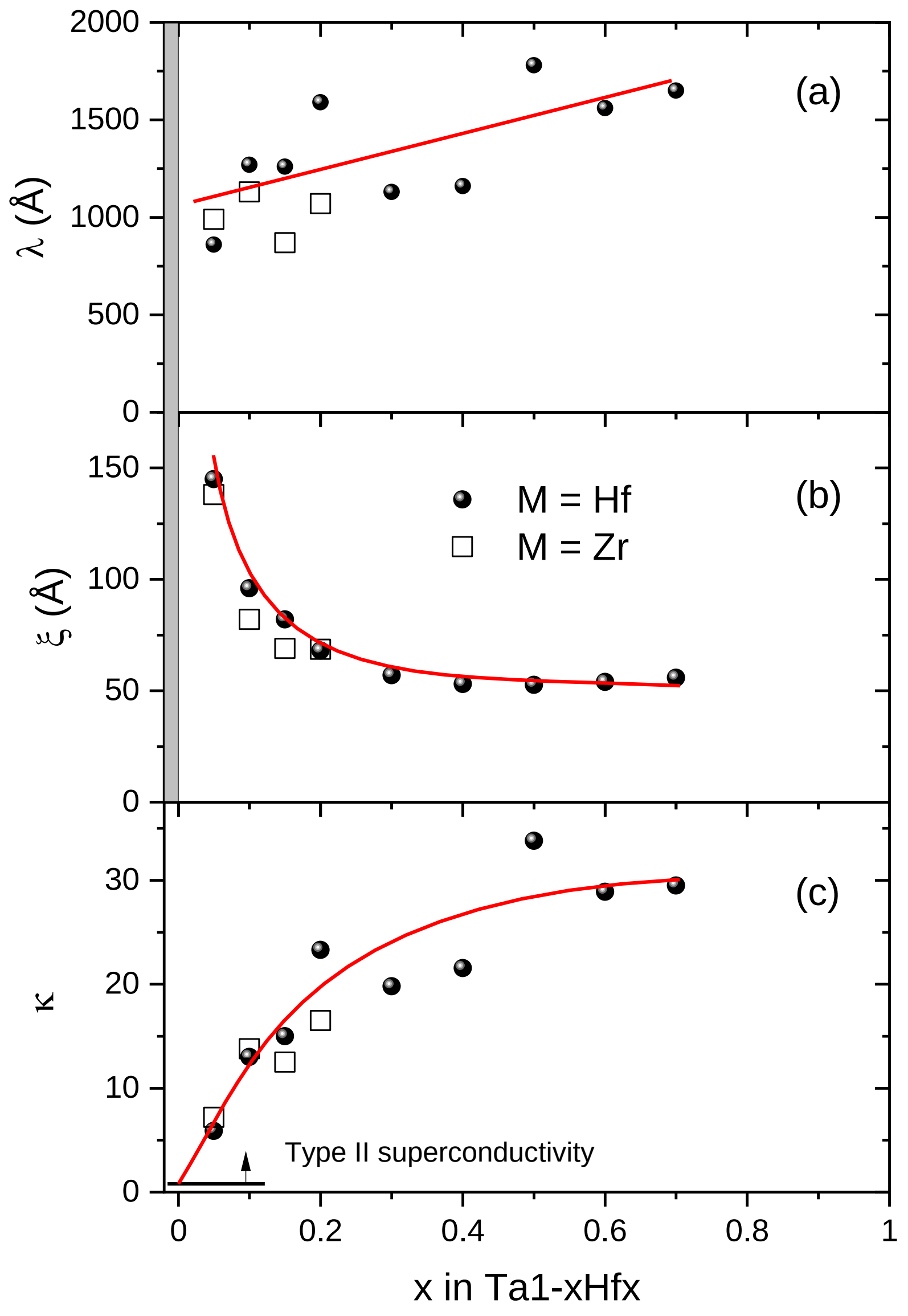

**Figure 8** Zero-field temperature dependence of the heat capacity $C_p/T$ for (a) Ta (open circles), (c) $Ta_{0.95}Hf_{0.05}$ (closed circles) and (e) $Ta_{0.95}Zr_{0.05}$ (open squares). Panel (b), (d) and (f) shows $C_p/T(T)$ under magnetic field of 3 T which is above the upper critical field for each sample.

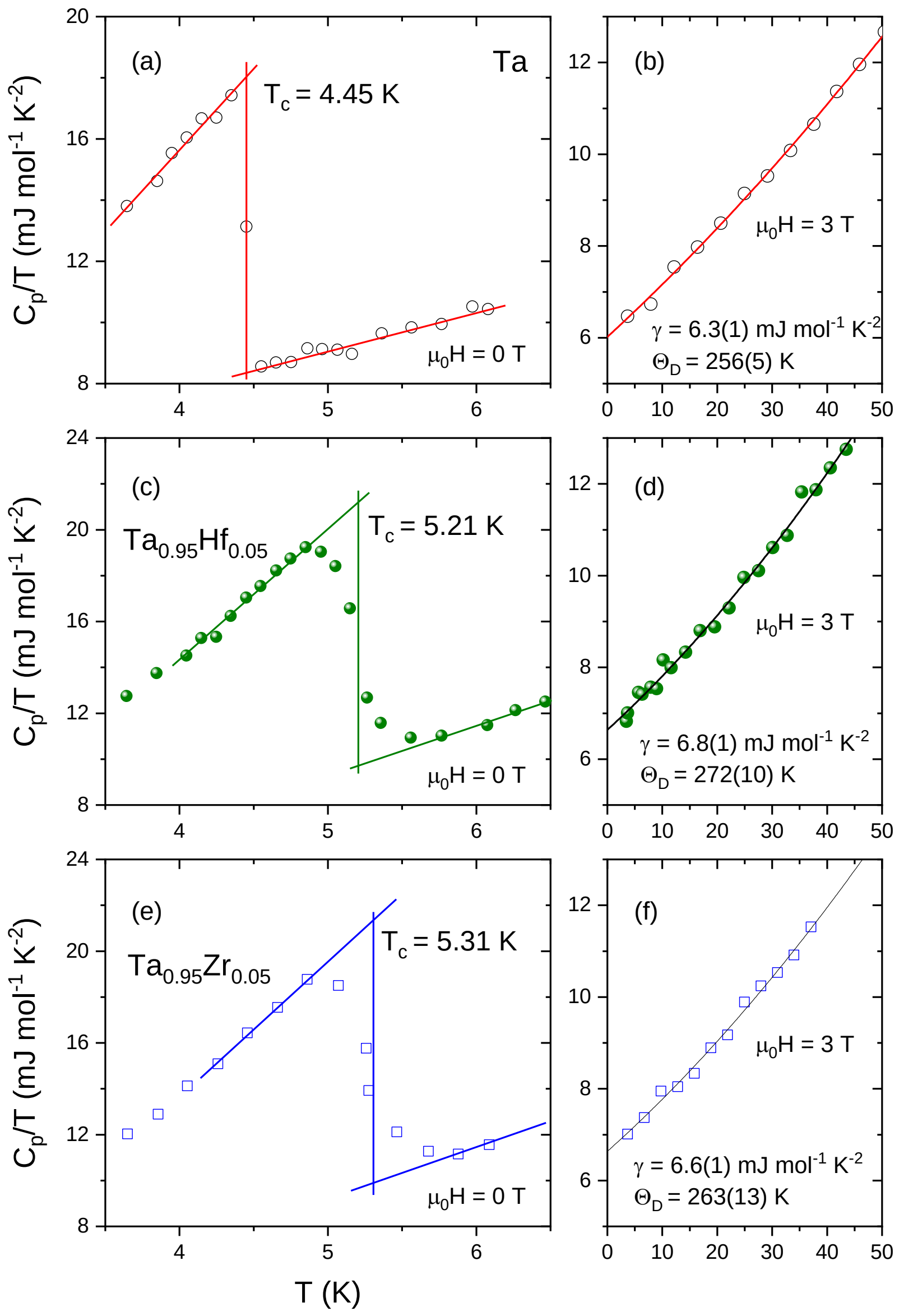

**Figure 9** (a) The Hf and Zr concentration dependence of the Sommerfeld coefficient (γ), (b) Debye temperature ($\Theta_D$) and (c) a normalized specific heat jump ($\Delta C_{p/}/\gamma T_c$). The red solid lines are guide lines.

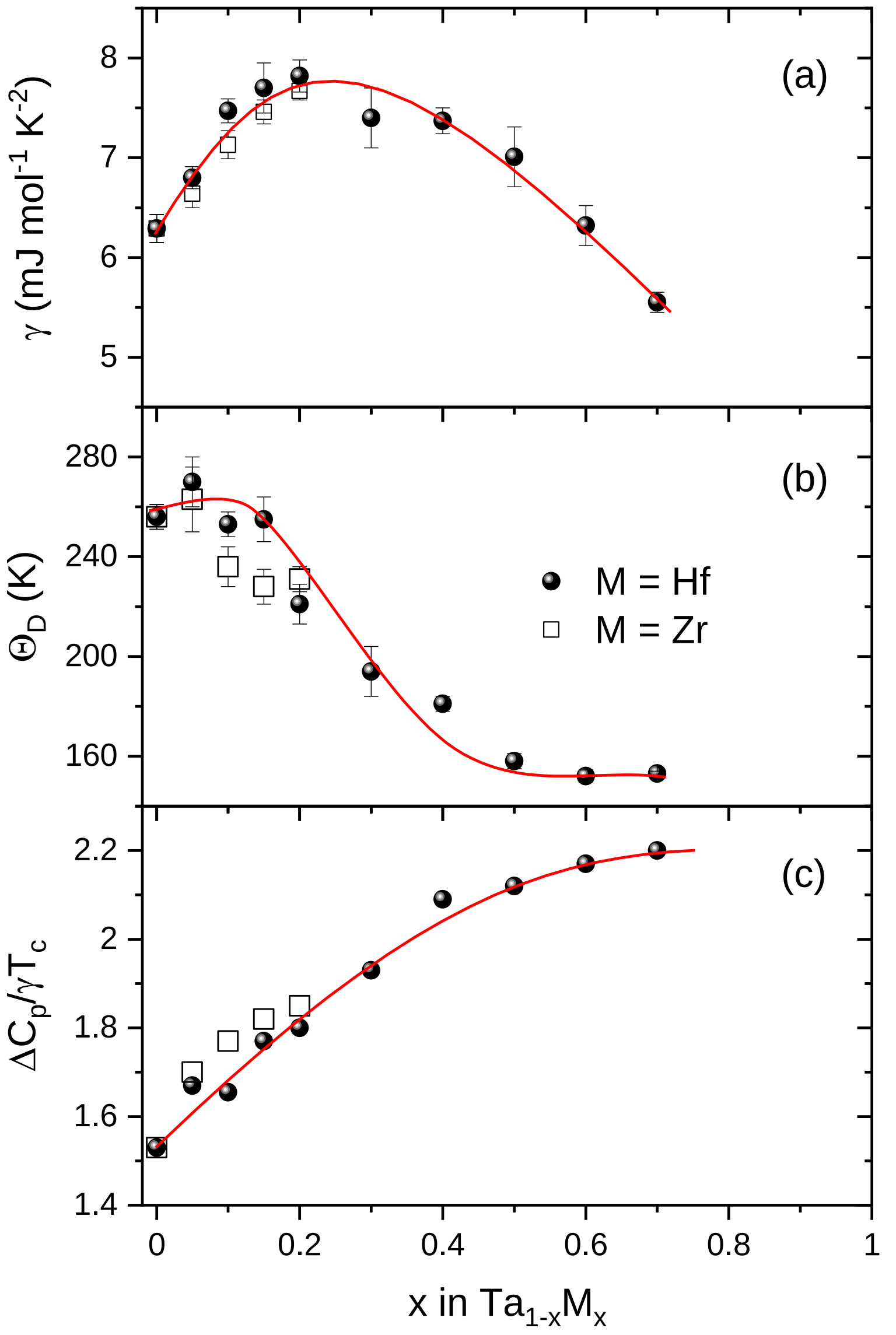

**Figure 10** Panel (a), (b) and (c) – the heat capacity $C_p/T$ vs. T in the vicinity of the superconducting transition under magnetic field: 0 Oe, 40 Oe, 80 Oe, 120 Oe and 160 Oe. The experimental data points were obtained by using a large, single heat pulse (ΔT 80% of the base temperature) and then re-processed by using a single-slope post processing method. Panel (d), (e) and (f) show $C_p/T$ H-T phase diagrams for Ta, $Ta_{0.95}Hf_{0.05}$ and $Ta_{0.95}Zr_{0.05}$, respectively.

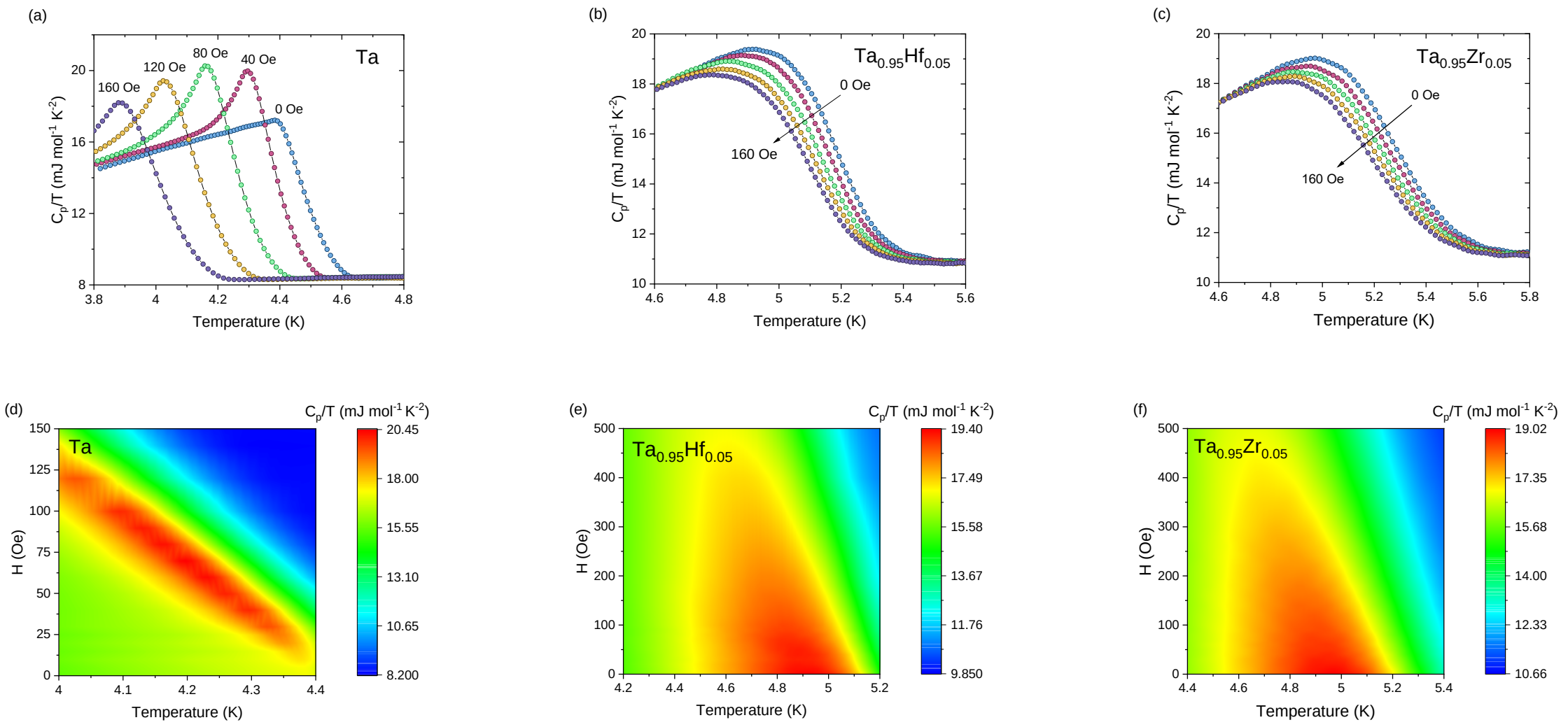

**Figure 11** Valence electron count ($N_{el}$/at.) dependence of (a) the logarithmically averaged phonon frequency ($\omega_{ln}$), (b) electron-phonon coupling constant, (c) electron density of states at the Fermi energy (DOS($E_F$)) and (d) superconducting critical temperature ($T_c$). The red solid lines are guide lines. The top axis shows the Hf or Zr concentration (x) in $Ta_{1-x}M_x$.

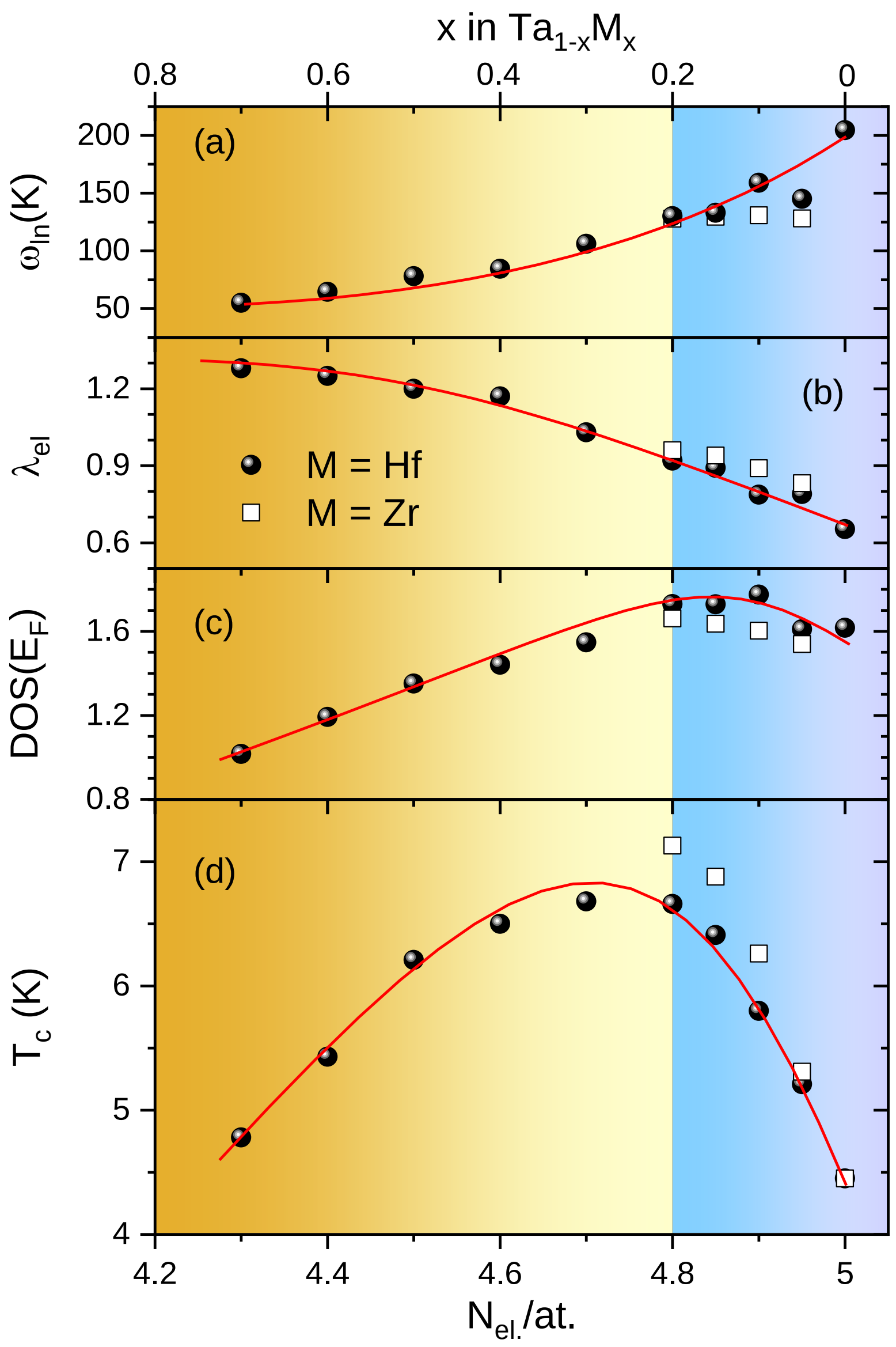

**Figure S1** Relative ration of the ZFC and FC magnetic susceptibility signal for $Ta_{1-x}Hf_x$. Data were collected at 1.8 K under field 10 Oe.

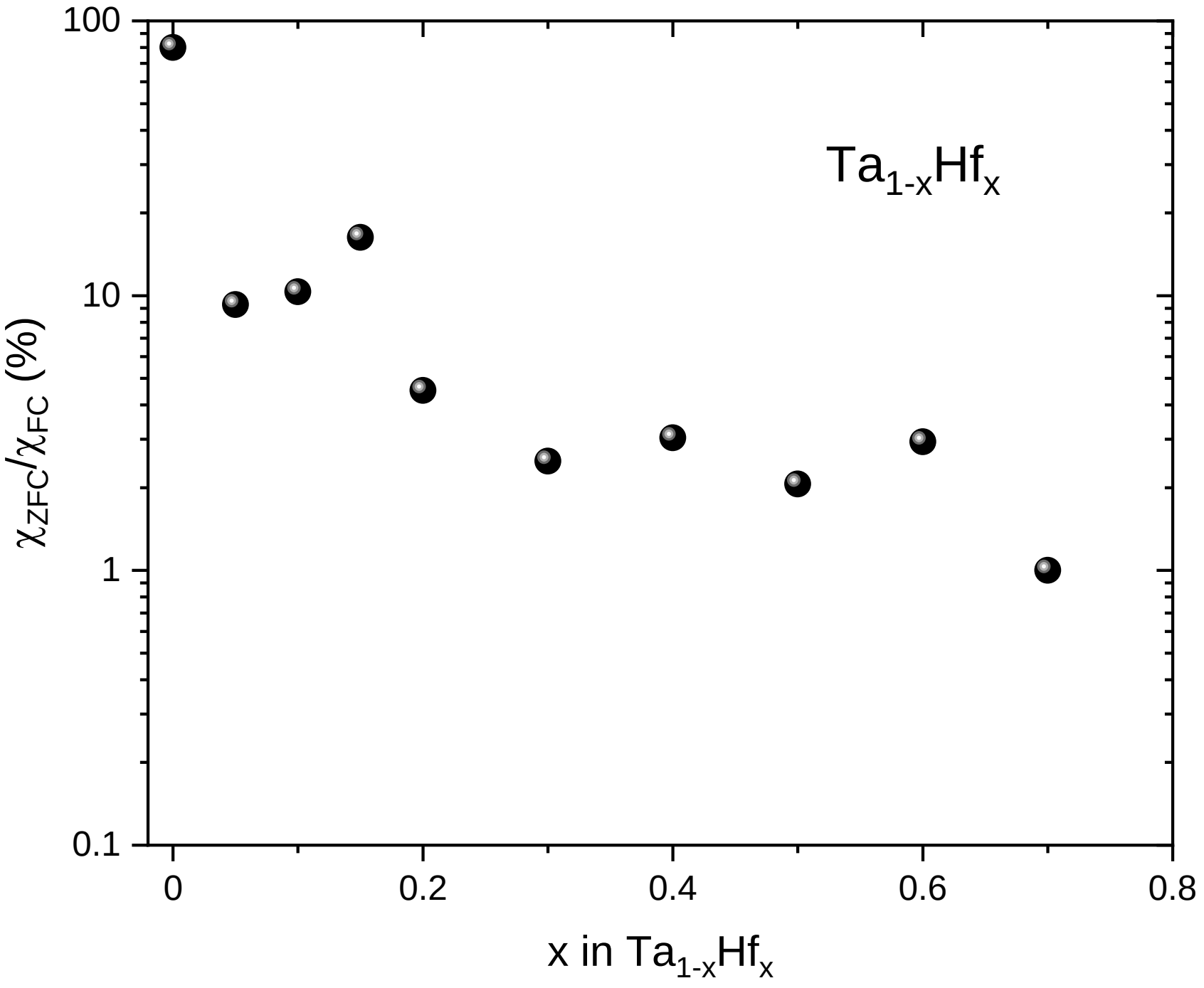

**Figure S2** Temperature dependence of the normalized electrical resistivity ρ(T)/ρ(300K) for selected $Ta_{1-x}Zr_x$ samples (x = 0, 0.05, 0.1, 015 and 0.2).

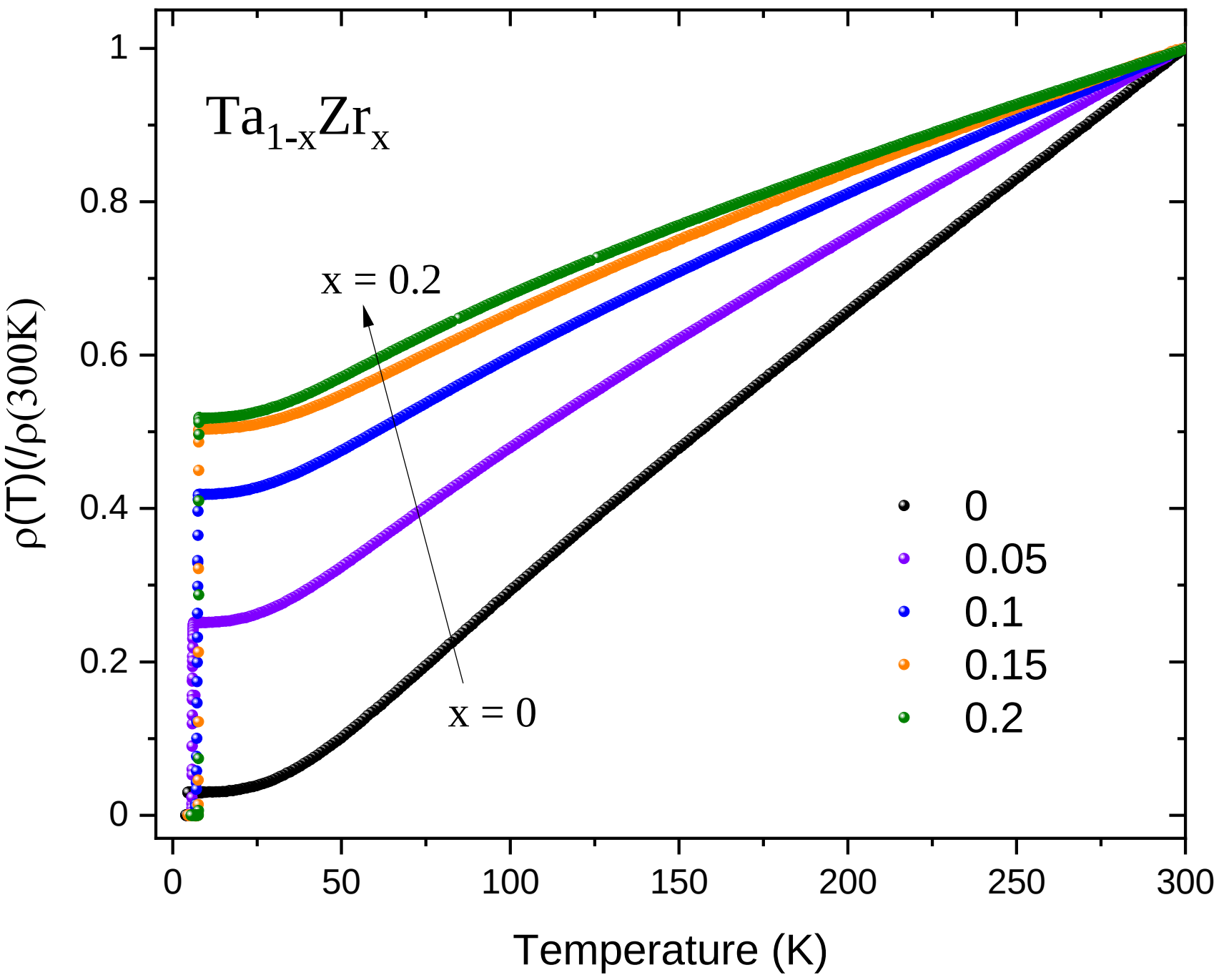

**Figure S3** The heat capacity $C_p/T$ vs. T for Ta (panel (a)) and Nb (panel (b)) in the vicinity of the superconducting transition measured under magnetic low magnetic field. The experimental data points were obtained by using a large, single heat pulse ($\Delta$T 80% of the base temperature) and then re-processed by using a single-slope post processing method. Panels (c) and (d) represent $C_p/T$ H-T phase diagrams for pure Ta (c) and pure Nb (d) metals. Clear difference between type I (Zr) and type II (Nb) superconductivity is observed.

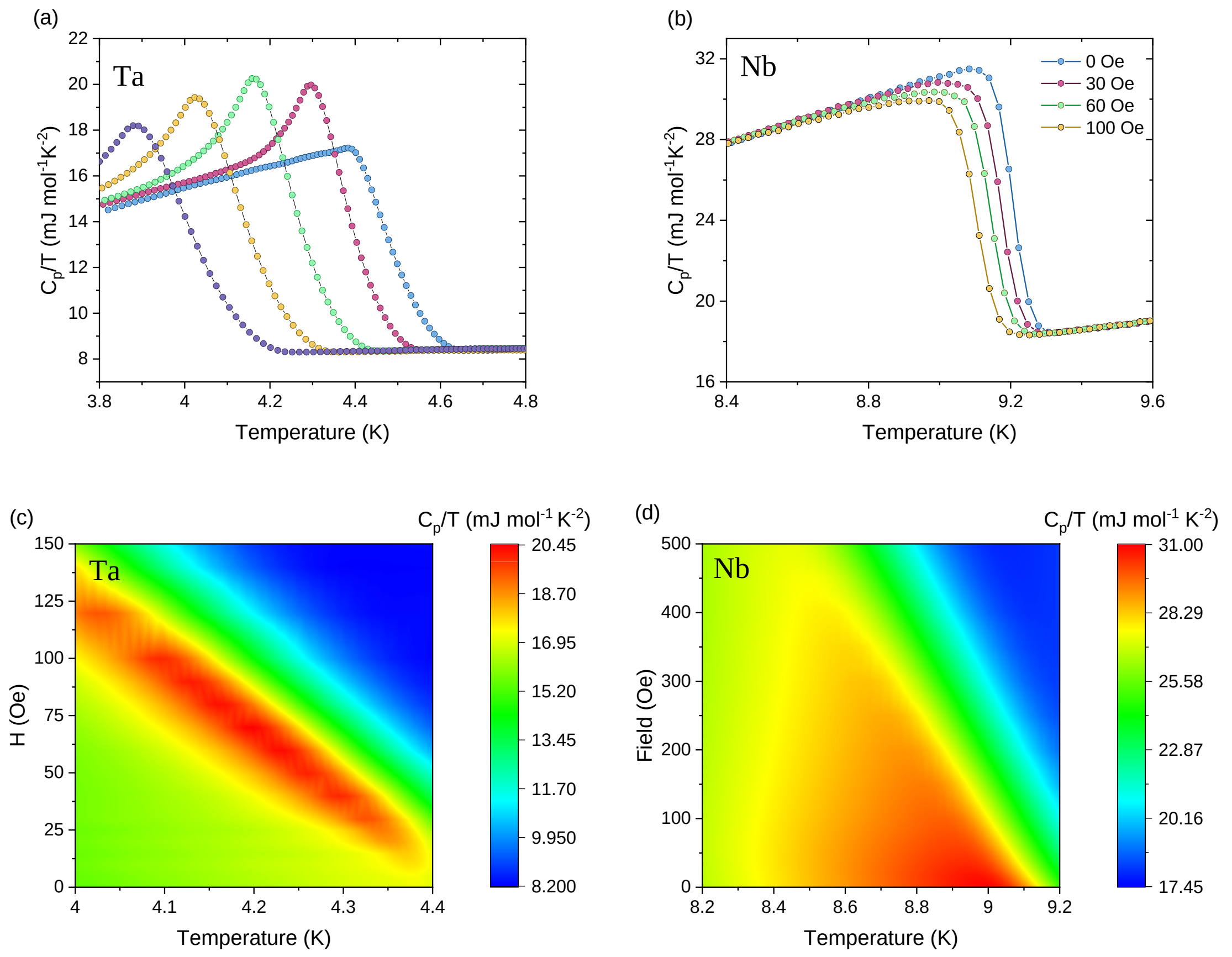